\documentclass[aps, 10pt, final, notitlepage, oneside, twocolumn, nobibnotes,
nofootinbib,superscriptaddress, noshowpacs,amsmath,amssymb]
{revtex4}
\usepackage{bm}
\usepackage{epsfig}
\usepackage{graphics}

\usepackage[english]{babel}
\begin{document}

\title{Electrocrystallization of Supercooled Water Confined between Graphene Layers}

\author{\firstname{Ramil~M.}~\surname{Khusnutdinoff}}\email{khrm@mail.ru} \affiliation{Kazan Federal University, 420008 Kazan, Russia}
\affiliation{Udmurt Federal Research Center, Siberian Branch, Russian Academy of Sciences, Izhevsk, 426068 Russia}

\author{\firstname{Anatolii~V.}~\surname{Mokshin}}\email{anatolii.mokshin@mail.ru}\affiliation{Kazan Federal University, 420008 Kazan, Russia}
\affiliation{Udmurt Federal Research Center, Siberian Branch, Russian Academy of Sciences, Izhevsk, 426068 Russia}

\begin{abstract}
A key feature of the crystallization of supercooled water confined in an applied static electric field is that the structural order here is determined not only by usual thermodynamic and kinematic factors (degree of supercooling, difference between chemical potentials for a liquid and a crystal, and viscosity) but also by the strength and direction of the applied electric field, size of a system (size effects), and the geometry of bounding surfaces. In this work, the electrocrystallization of supercooled water confined between ideally flat parallel graphene sheets at a temperature of $T=268$~K has been considered in this work. It has been established that structural order is determined by two characteristic modes. The initial mode correlates with the orientation of dipolar water molecules by the applied electric field. The subsequent mode is characterized by the relaxation of a metastable system to a crystalline phase. The uniform electric field applied perpendicularly to the graphene sheets suppresses structural ordering, whereas the field applied in the lateral direction promotes cubic ice $I_c$.
\end{abstract}
\maketitle

Although water is one of the most widespread substances on the Earth and is studied in numerous experimental, theoretical, and numerical works, many of its physical properties are still poorly studied \cite{Gallo2016,Debenedetti2003,Brazhkin2015,Fomin2015,Khusnutdinoff2011,Khusnutdinoff2012}. Owing to the specific arrangement of two hydrogen atoms with respect to the oxygen atom, the electron density distribution in the water molecule is quite nonuniform. In spite of the presence of distinguished directions in the interaction between water molecules, which can be attributed to some well-known crystalline phases typical of single-component systems, water has a fairly complex phase diagram, which contains numerous stable and metastable phases, several triple points, and one or possibly even two critical points. In particular, the authors of \cite{Stanley1994,Kanno2006,Ferreira2015,Stanley2019} actively discussed the possibility of the existence of the second critical point in water\footnote{We note that the existence of the second critical point for water was demonstrated in molecular dynamics simulations with various model potentials \cite{Paschek2005,Jedlovszky2005,Abascal2011}. Similar results were obtained for a number of other tetrahedrally coordinated liquids \cite{Glosli1999,Lacks2000,Poole2001}.}. It has been reliably established that the phase diagram of water contains at least 17 crystalline phases \cite{Loerting2009,Loerting2006,Falenty2014,Malenkov2006} and three amorphous phases, including low-density amorphous ice, high-density amorphous ice, and very high density amorphous ice \cite{Brazhkin1999,Giovambattista2005,Mishima2013}. 

Studies of the properties of water confined in limited spatial domains with specific geometries have recently become of particular importance. For example, this concerns water located in cylindrical pores in solids or between thin films or plates \cite{Grigorieva2015}. When the size of the bounding region is comparable with the characteristic range of the interaction between water molecules, the so-called size effects begin to affect many physical properties of water. The phase diagram of spatially confined water has a number of specific features: existence of metastable equilibrium phases (e.g., \textit{dodecagonal quasicrystals} \cite{Kastelowitz2010}), shift of the melting line, and change in the domains of equilibrium phases \cite{Erko2011,Bai2012,Khusnutdinoff2013}. The inclusion of size effects is usually a nontrivial task. It is remarkable that just these effects are responsible for various processes involving water, e.g., diffusion in cell membranes and water transport in pores of plants. 

Particular attention is focused on the study of the effect of external fields (electric, magnetic, ultrasonic, etc.) on the crystallization of confined water \cite{Qian2014,Zhu2014,Terrones2016,Khusnutdinoff2019}. The crystallization of water in an applied electric field —- \textit{electrocrystallization} -- is a key phenomenon for
such processes as the formation of atmospheric precipitation, formation and growth of ice on the wings of aircraft, and cryoconservation of biological materials \cite{Lumsdon2004,Alexander2019}.

The aim of this work is to study the effect of the strength and direction of a uniform electric field on the process of crystallization of spatially confined supercooled water.
\begin{figure}[h]
	\centering
	\includegraphics[width=8cm]{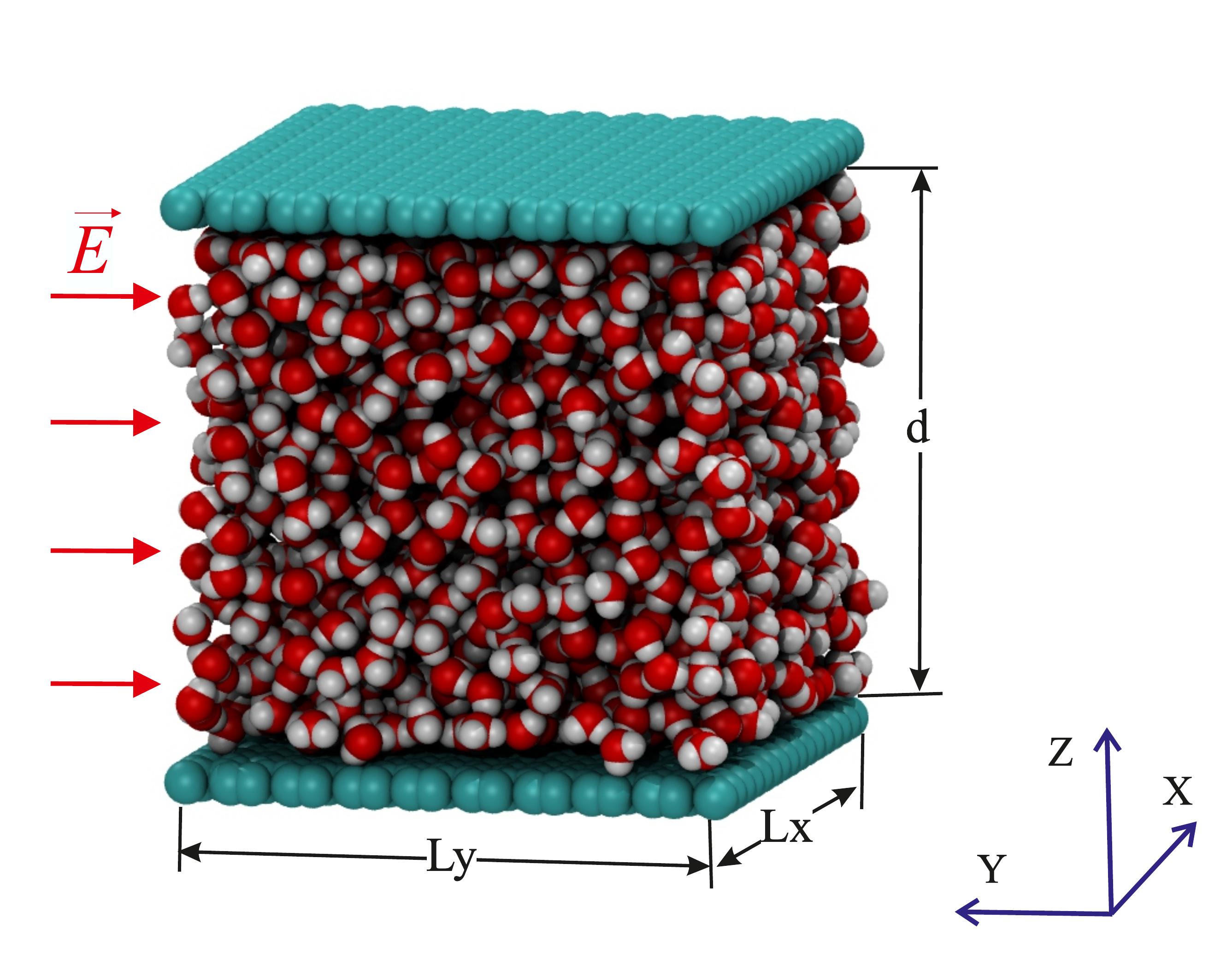}
	\caption{(Color online) Schematic of the simulated system at a temperature of $T=268$~K and a density of $\rho=0.92$~g/cm$^3$.}
	\label{Fig_01}
\end{figure}
We consider a supercooled water layer confined between two parallel infinite graphite
films (see Fig.~\ref{Fig_01}). The graphene sheets are spaced by about ten average linear dimensions of the water molecule. The interaction between water molecules is
described by means of the Tip4p/Ice atomistic model potential \cite{Abascal2005}, which correctly reproduces the phase diagram \cite{Vega2011}. Intramolecular bonds are taken into account using the SHAKE algorithm \cite{Ryckaert1977}. The energy of electrostatic interactions is calculated using the PPPM method with the cutoff radius $r_c=\textrm{13~\AA}$ \cite{Rajagopal1994}. The interaction between carbon atoms of graphene sheets with water molecules is described by a Lennard-Jones potential whose parameters are determined using the Lorentz-Berthelot mixing rule \cite{Gordillo2010,Khusnutdinoff2014}. The simulation is performed in an isothermal-isochoric $(NVT-)$ ensemble at a temperature of $T=268$~K for
densities in the range $\rho\in[0.90; \; 0.98]$~g/cm$^3$. To stabilize the temperature, we use the Nose-Hoover thermostat algorithm with a relaxation parameter of $1.0$~ps \cite{Nose1984,Hoover1985}. The supercooled water phase at a temperature
of $T=268$~K was obtained from high-temperature liquid water at $T=350$~K by rapid cooling at a rate of $\gamma\simeq1.0$~K/ps \cite{Mokshin2017}. Periodic boundary conditions are imposed along the \textit{OX} and \textit{OY} directions. The external
uniform static electric field with the strength  $E=0.5~\textrm{V/\AA}$ can be applied either along the \textit{Z} axis perpendicular to the graphite layers or along the \textit{Y} axis parallel to the layers. This strength of the electric field corresponds to those acting on water molecules near the surfaces of biopolymers \cite{Drost-Hansen} and in cracks of amino acids \cite{Gavish1992}. According to \textit{ab-initio} molecular dynamics simulations, electric fields above the threshold value $E_c\sim 0.35~\textrm{V/\AA}$ can dissociate bulk water with the formation of a stable ion current \cite{Saitta2012}, whereas confined
water dissociates at higher electric field strengths $E>0.5~\textrm{V/\AA}$ \cite{Qiu2013}.
\begin{figure}[h]
	\centering
	\includegraphics[width=11cm]{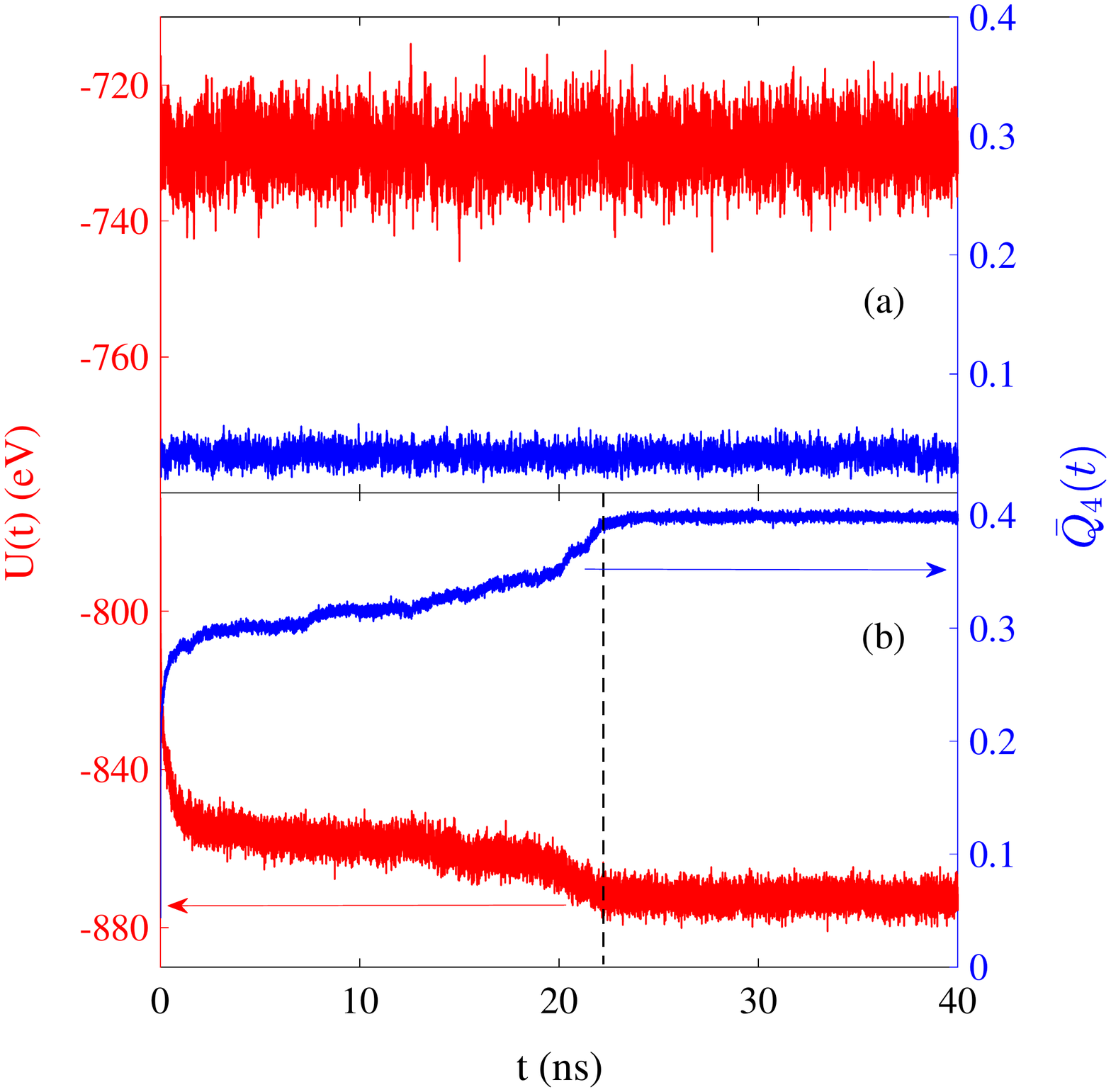}
	\caption{(Color online) Time dependences of the (red line) internal energy and (blue line) global orientational order parameter $\bar{Q}_4$ of supercooled confined water at a temperature of $T=268$~K, density of $\rho=0.92$~g/cm$^3$, and external electric field strength of $E=0.5~\textrm{V/\AA}$ applied along the (a) \textit{Z} axis perpendicular to the graphite layers and (b) \textit{Y} axis along the graphite layers.}
	\label{Fig_02}
\end{figure}

\begin{figure*}[t]
	\centering
	\includegraphics[width=17cm]{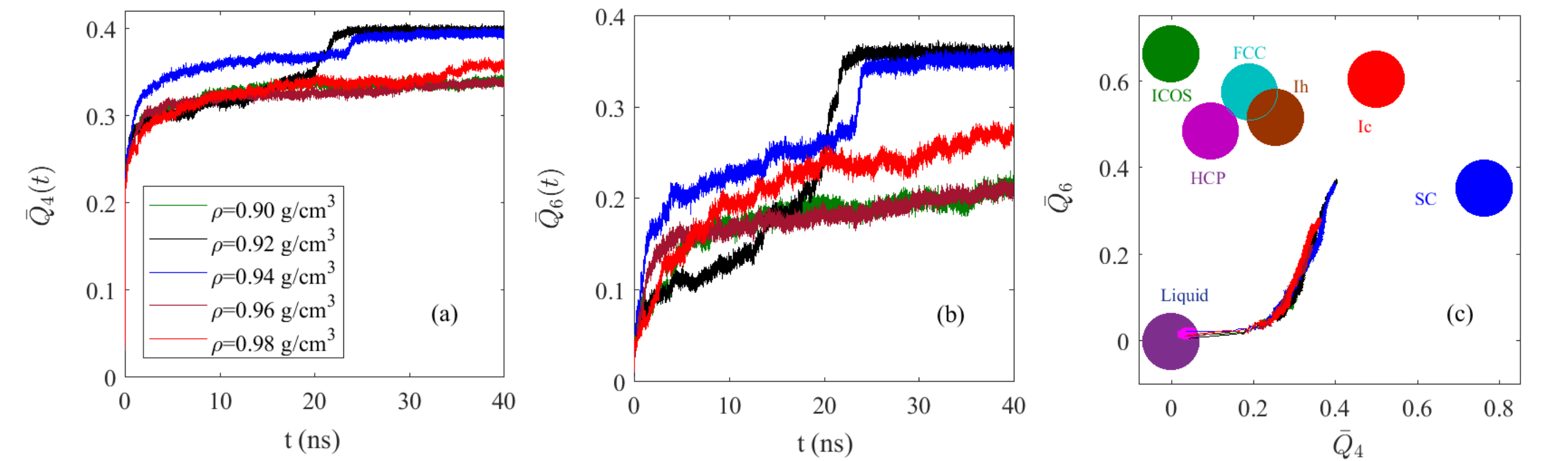}
	\caption{(Color online) Time dependences of the global order parameters  (a) $\bar{Q}_4$ and (b) $\bar{Q}_6$ for confined water in the external electric field $E=0.5~\textrm{V/\AA}$ at a temperature of $T=268$~K and various densities; (c) trajectories of the formation of the order parameter	on the ($\bar{Q}_4$,$\bar{Q}_6$) diagram with the regions of (Ih) hexagonal ice and (Ic) cubic ice and (SC) simple cubic, (FCC) face-centered cubic, (HCP) hexagonal close-packed, and (ICOS) icosahedral ideal crystal structures.}
	\label{Fig_03}
\end{figure*}
To reveal the effect of the electric field on the process of structural ordering in supercooled water, we use the structure analysis based on the calculation of
the global Steinhardt-Nelson-Ronchetti orientational order parameter~\cite{Steinhardt1983,Lechner2008}
\begin{equation}
\bar{Q}_{l}=\left(\frac{4\pi}{2l+1}\sum_{m=-l}^{l}\left|\frac{\sum_{i=1}^{N}\sum_{j=1}^{n_{b}^{(i)}}Y_{lm}(\theta_{ij},\varphi_{ij})}{\sum_{i=1}^{N}n_{b}^{(i)}}\right|^{2}\right)^{1/2}.\nonumber
\end{equation}
Here, $n_{b}^{(i)}$ is the number of nearest neighbors for the $i$th molecule, $Y_{lm}(\theta_{ij},\varphi_{ij})$ are the spherical harmonics, $\theta_{ij}$ and $\varphi_{ij}$ are the polar and azimuth angles, respectively. Each type of crystal lattice is characterized by a unique set of orientational order parameters $\bar{Q}_{l}$ (where $l = 4, 6, 8,...$). This makes it possible to identify the presence of an ordered structure of a certain type in the system under consideration. In particular, $\bar{Q}_4=0.259$ and $0.506$ for ideal hexagonal and cubic ices, respectively. The parameters $\bar{Q}_4$ and $\bar{Q}_6$ for disordered systems (water amorphous ice) are close to zero~\cite{Galimzyanov2019}.

Figure~\ref{Fig_02} shows the time dependences of two order parameters - the internal energy $U$ and orientational order parameter $\bar{Q}_4$ -- calculated for confined water crystallized at a temperature of $T=268$~K and a density of $\rho=0.92$~g/cm$^3$ in the electric field with the strength $E=0.5~\textrm{V/\AA}$ applied along (a) the $Z$ axis perpendicular to the graphite layers and (b) the $Y$ axis along the layers. According to Fig.~\ref{Fig_02}, both parameters are identically informative for detection of structural changes in water. In particular, when the electric field is applied along the \textit{OZ} direction perpendicular to
(graphite layers, no structural transformations occur in a time of $40$~ns (see Fig.~\ref{Fig_02}a). When the electric field is directed along the system (\textit{OY} direction), a fairly fast crystallization of water occurs, which is clearly seen on the time dependences $U(t)$ and $\bar{Q}_4(t)$. The internal energy decreases to $U=-872$~eV and the orientational order parameter increases to $\bar{Q}_4=0.403$, which is characteristic of an ordered system (Fig. \ref{Fig_02}b). The time scale of the electrocrystallization of the sample with the characteristics presented in Fig.~\ref{Fig_02}b is $\tau\sim 22$~ns. The observed dependence of structural changes on the orientation of the electric field is due to the presence of confining surfaces in the \textit{OZ} direction, which prevent reaching lower densities typical of crystalline phases of water at the orientation of dipoles along the \textit{OZ} direction.

Figures~\ref{Fig_03}a and \ref{Fig_03}b show the time dependences of the global order parameters (a) $\bar{Q}_4$ and (b) $\bar{Q}_6$ for confined water with different densities subjected to the external electric field $E=0.5~\textrm{V/\AA}$ applied along the \textit{OY} direction. The plotted noisy curves $\bar{Q}_4(t)$ and  $\bar{Q}_6(t)$ clearly demonstrate two structural ordering modes. The first mode is due to the orientation of dipole water molecules by the electric field, leading to the formation of a quasiordered structure with the characteristic order parameters $\bar{Q}_4=0.34\div0.4$ and $\bar{Q}_6=0.2\div0.28$.
The second mode determines further ordering and is associated with usual relaxation to the equilibrium crystalline phase. In this mode, $\bar{Q}_4=0.399$ and $\bar{Q}_6=0.36$. Both modes should be characterized by time scales depending on the density of the system and the external electric field strength $E$. These time scales cannot be estimated from the results of numerical experiments for a few densities shown in Figs.~\ref{Fig_03}a and \ref{Fig_03}b.

\begin{figure}[h]
	\centering
	\includegraphics[width=8cm]{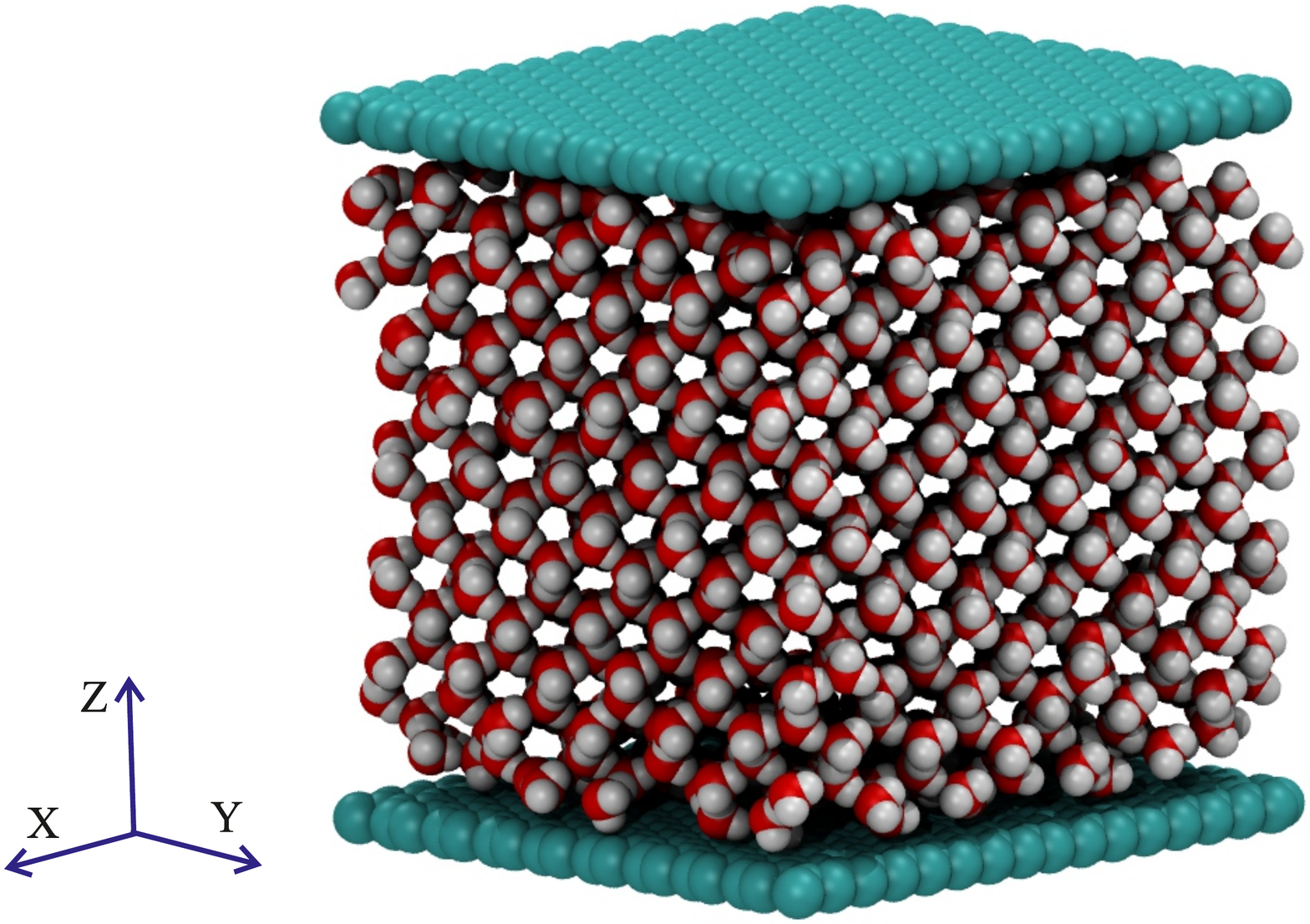}
	\caption{(Color online) Configuration of the system at a
		temperature of $T=268$~K and density of $\rho=0.92$~g/cm$^3$ in
		the external electric field with the strength $E=0.5~\textrm{V/\AA}$.}
	\label{Fig_04}
\end{figure}

Figure~\ref{Fig_03}c shows the trajectories of the formation of the ordered phase on the ($\bar{Q}_4$,$\bar{Q}_6$) diagram. It is remarkable that the crystallization of water with different indicated densities is characterized on the ($\bar{Q}_4$,$\bar{Q}_6$) diagram by a common trajectory tending to cubic ice Ic. The order parameters $\bar{Q}_4$ and $\bar{Q}_6$ are maximal at the densities $\rho=0.92$ and $0.94$~g/cm$^3$, respectively. According to the quantum mechanical calculations \cite{Geiger2014} and experimental data \cite{Brazhkin2003,Murray2005}, cubic ice, first, should be unstable in a bulk macrosystem and, second, appears at lower temperatures. In our situation, the cubic ice structure is also identified by the observed hexagonal channels (water molecules form a diamond-like packing of ice Ic in the (111) plane) characteristic of this ice type (see Fig.~\ref{Fig_04}). Thus, the application of the external electric field results in the formation of cubic ice Ic instead of the most widespread hexagonal ice Ih. It is noteworthy that cubic ice is thermodynamically less stable than ordinary hexagonal ice (except for a narrow low-temperature range of $123-153$~K), although their structures are very similar. Each water molecule forms four hydrogen bonds and satisfies the so-called ``ice rules'' in both phases \cite{Wang2018}.
The distribution of molecules in the hexagonal and cubic ice phases has a nearly perfect tetrahedral symmetry. The (001) basal plane of ice Ih is identical to the
(111) plane of ice Ic. 
It is remarkable that the results obtained are in agreement with the conclusions made in \cite{Grigorieva2015}, where the high-resolution electron microscopy study of a water nanofilm confined between two graphene sheets revealed a quadratic ice lattice, which has a symmetry qualitatively differing from the ordinary tetrahedral geometry of hydrogen bonds between water molecules. This type of ice can be considered as a cubic ice layer in the (110) plane, which is characterized by a high packing density with a lattice constant of 2.83~\rm{\AA}. Such an ice can be formed inside hydrophobic nanochannels in the form of two- and three-layer crystallites.

The features of the crystallization of the system under consideration are determined by three circumstances. The first circumstance is associated with the usual relaxation of supercooled water to the crystalline phase. The second circumstance is the effect of the specific geometry of the system and the presence of two graphene planes bounding the system on crystallization. Finally, the third circumstance conerns the influence of the magnitude and direction of an external electric field on the system. These three circumstances determine the characteristic time (rate) of
crystallization and the final crystalline phase. In particular, an increase in the size of the system suppresses the effect of the external field on the crystallization of water. Furthermore, the authors of \cite{Stan2011} reported that
the external electric field only slightly affects the rate of homogeneous crystal nucleation of ice in supercooled \textit{millimeter-sized} water droplets. At the same time, the crystal nucleation rate and the size of the critical nucleus in \textit{bulk} water are determined only by the degree of supercooling and are independent of the field strength \cite{Zaragoza2018}. It is remarkable that the melting temperature of water increases with the field strength. In particular, \textit{ferroelectric cubic ice} $I_{cf}$ in an electric field of about several volts per angstrom nucleates and grows under the same thermodynamic conditions as
hexagonal ice Ih in zero field. To conclude, we note that our results are in agreement with the conclusion made in \cite{Johari2005} that cubic ice is the most favorable crystalline structure for water confined in pores or thin
films with a thickness up to $100~\textrm{\AA}$.

\section{ACKNOWLEDGMENTS}
Extensive molecular dynamics simulations were performed using the equipment of the Supercomputer Center, Moscow State University, and the Computational Cluster,
Kazan Federal University.

\section{FUNDING}
This work was supported by the Russian Science Foundation (project no. 19-12-00022).

\vfill\eject
\end{document}